\documentclass[utf8]{frontiersSCNS} 

\usepackage{url,hyperref,microtype,subcaption}
\usepackage[onehalfspacing]{setspace}

\def\keyFont{\fontsize{8}{11}\helveticabold }
\def\firstAuthorLast{Zenteno-Quinteros \& Moya} 
\def\Authors{Bea Zenteno-Quinteros\,$^{1,*}$, Pablo S. Moya\,$^{1,*}$}
\def\Address{$^{1}$Departamento de F\'isica, Facultad de Ciencias, Universidad de Chile, Santiago, Chile }
\def\corrAuthor{Bea Zenteno-Quinteros}
\def\corrEmail{beatriz.zenteno@ug.uchile.cl}

\begin{document}
\onecolumn
\firstpage{1}

\title[core-strahlo heat-flux instability]{The role of core and strahlo electrons properties on the whistler heat-flux instability thresholds in the solar wind} 

\author[\firstAuthorLast ]{\Authors} 
\address{} 
\correspondance{} 

\extraAuth{Pablo S. Moya\\pablo.moya@uchile.cl}

\maketitle

\begin{abstract}
There is wide observational evidence that electron velocity distribution functions (eVDF) observed in the solar wind generally present enhanced tails and field-aligned skewness. These properties may induce the excitation of electromagnetic perturbations through the whistler heat-flux instability (WHFI), that may contribute to a non-collisional regulation of the electron heat-flux values observed in the solar wind via wave-particle interactions. Recently, a new way to model the solar wind eVDF has been proposed: the core-strahlo model. This representation consist in a bi-Maxwellian core plus a Skew-Kappa distribution, representing the halo and strahl electrons as a single skewed distribution. The core-strahlo model is able to reproduce the main features of the eVDF in the solar wind (thermal core, enhanced tails, and skewness), with the advantage that the asymmetry is controlled by only one parameter. In this work we use linear kinetic theory to analyze the effect of solar wind electrons described by the core-strahlo model, over the excitation of the parallel propagating WHFI. We use parameters relevant to the solar wind and focus our attention on the effect on the linear stability introduced by different values of the core-to-strahlo density and temperature ratios, which are known to vary throughout the Heliosphere. We also obtain the stability threshold for this instability as a function of the electron beta and the skewness parameter, which is a better indicator of instability than the heat-flux macroscopic moment, and present a threshold conditions for the instability that can be compared with observational data.

\tiny
 \keyFont{ \section{Keywords:} solar wind, heat-flux, electron instabilities, Skew-Kappa distributions} 
\end{abstract}

\section{Introduction}

Recent observations have shown that electron heat-flux measurements in the solar wind are not completely explained by the collisional transport model given by the Spitzer-Härm law \citep{spitzer1953transport}. The field aligned electron heat-flux at 1~AU from the Sun is consistent with this model only up to a Knudsen number $K_n \sim 0.3$, where $K_n$ is the ratio between the mean free path and the temperature gradient scale. Beyond that, the observed heat-flux values are lower than those predicted by this law \citep{bale2013electron}, which suggest that there exist non-collisional processes relevant to fully understand the electron thermal energy transport in the Heliosphere. Moreover, data also suggest that non-collisional mechanisms, e.g. electron micro-instabilities, may play an important role in the near-Sun environment as the heat-flux observations do not follow the Spitzer-Härm law for any range of the estimated $K_n$~\citep{halekas2021electron}. 

Electron heat-flux instabilities (HFI) in the solar wind are wave modes excited by the free energy provided by the skewness of the electron velocity distribution function (eVDF) along to the interplanetary magnetic field \citep{gary1999electron, shaaban2018beaming, lopez2020alternative}. Among other non-thermal features of the eVDF, this field-aligned skewness is clearly observed in solar wind's in-situ measurements \citep{feldman1975solar, pilipp1987variations, nieves2008solar, stverak2009radial}. Considering that the electron heat-flux is closely related to the eVDF skewness, the HFIs are the main candidates to be the non-collisional mechanism that self-regulate the heat-flux values in the solar wind, via wave-particle interactions. Therefore, they may explain the observed electron heat-flux profile in the solar wind \citep{gary1975heat, shaaban2018clarifying, lopez2019particle}. Among these skewness-driven instabilities, the excitation of the whistler mode of the electron cyclotron branch, known as the whistler heat-flux instability (WHFI), has been often invoked as one of the most probable non collisional processes regulating the electron heat-flux \citep{gary1994whistler, gary2000whistler, kuzichev2019nonlinear, shaaban2019quasi}. However, the dominant wave mode is still under debate and recent works even suggest that it may not be possible to identify only one instability as the principal non-collisional mechanism \citep{lopez2020alternative}. Thus, studies regarding the electron heat-flux regulation in the solar wind should consider the interplay and/or succession of different instabilities \citep{shaaban2019interplay}.

Different theoretical and observational studies have tried to assess the importance of these HFI on the non-collisional regulation of the electron heat-flux in the solar wind. From the observational point of view, these studies focus on comparing measurements of the normalized electron heat-flux macroscopic moment in the solar wind with analytical expressions of marginal stability thresholds of electron HFI \citep{gary1999electron, bale2013electron, tong2019statistical, halekas2021electron, cattell2022parker}. In \citet{bale2013electron} the authors contrast data obtained by the WIND spacecraft with theoretical thresholds values for the whistler and magnetosonic instabilities. They conclude that for the data set analyzed, the WHFI over constrain the observations, and the magnetosonic instability is more consistent in the collisionless regime when $K_n>0.3$ and the plasma beta is large. In addition, in \citet{halekas2021electron} the authors used data provided by the Parker Solar Probe at heliocentric distances between 0.125 and 0.25 AU from the Sun, and concluded that the observed heat-flux dependence on plasma beta is consistent with theoretical thresholds associated with oblique whistler waves generated via the fan instability~\cite{vasko2019whistler}. In contrast, in \citet{cattell2022parker}, authors showed that whistlers waves are extremely rare inside $\sim$0.13 AU and the heat-flux vs beta relationship is not constrained by the heat-flux fan instability this close to the Sun. 

Theoretical linear and quasilinear approximations, as well as particle simulations, have been used to address this issue. However, to develop these types of studies, it is necessary to model the eVDF. In the solar wind, the observed eVDF has been typically characterized in terms of three subpopulations: a quasithermal core at lower energies, which has most of the electron density; the suprathermal halo representing the enhanced high energy tails observed in the eVDF; and also the strahl, a suprathermal field-aligned beam which gives the eVDF its skewness. Under this context, different models have emerged to describe the plasma physics of solar wind electrons. Trying to mimic the observations, and to emulate the non-thermal characteristics of the electrons, most of the used models for the eVDF consists on the superposition of core, halo and/or strahl subpopulations. Among them, the most widely considered model consists on the superposition of two drifting bi-Maxwellian (typically core and strahl), which allows to have a skew distribution function \citep{gary1994whistler, gary2000whistler, kuzichev2019nonlinear, shaaban2019quasi, lopez2020alternative}. More realistic models have also been used to describe the eVDF in the solar wind, where Kappa distribution functions are considered to reproduce the high energy tails (the halo) of the observed eVDF \citep{saeed2016electron, shaaban2018clarifying, lazar2018temperature}. Furthermore, more exotic distributions have been also used to model the solar wind's suprathermal population, which by considering ad-hoc mathematical expressions are also able to address the electrons properties \citep{horaites2018,vasko2019whistler}.

Under this context, a new way to describe the electron population in the solar wind has recently been proposed by \citet{zenteno2021skew} (from now on paper A). In this work, the authors propose the so called ``core-strahlo model" as new way to describe the solar wind eVDF. This model consists on the sum of a drifting bi-Maxwellian (the core) and a Skew-Kappa function, representing halo and strahl in a single skew distribution. Therefore, using the superposition of only two functions, the model reproduces the 3 main kinetic features of the observed eVDF, namely: quasithermal core, enhanced tails and skewness. 
In paper A, the authors used the core-strahlo model and studied the effect of different plasma parameters on the excitation of the WHFI and its marginal stability thresholds. They showed that instead of the electron heat-flux moment $q_{e}$, which have been customarily used in to analyze the WHFI, the skewness parameter $\delta_s$ (i.e the parameter that controls the skewness of the core-strahlo distribution) is the most relevant when studying the WHFI. This is because high $\delta_s$ values rather than high $q_e$ values are consistent with more unstable states to the WHFI when a more realistic representation is used to model the eVDF in the solar wind. 

In paper A authors presented the core-strahlo model for the first time, and focused on the mathematical and technical details necessary to apply the model to the analysis of the whistler heat-flux instability (WHFI). They also compared the dispersion results with a two drifting Maxwellian model, and analyzed the instability as a function of the asymmetry parameter $\delta_s$, the kappa parameter and also plasma beta. To do so they fixed the density of the strahlo and also the core-to-strahlo temperature ratio. Along the same lines, in this work we expand the analysis performed in paper A. Here we use the core-strahlo model to describe the electron population and examine how the WHFI behaves as the strahlo-to-core temperature ratio and the strahlo number density are modified. Thus, here we complement and complete the systematic analysis of the instability as a function of all relevant parameters that was started with paper A. Additionally, following Peter Gary's legacy, we obtain the marginal stability thresholds and analyze how they change as we modify these parameters. Indeed, in situ measurements show that these two parameters exhibit several values as a function of heliocentric distance and solar wind speed \citep{maksimovic2005radial, lazar2020characteristics}. Thus, a systematic study on how the WHFI depends on density and temperature ratios of the solar subpopulations, becomes relevant for the understanding of the regulation of electron heat-flux in the solar wind. Accordingly, this article organized as follows: in Section 2 we briefly describe the core-strahlo model and its properties, and performed the stability analysis of the parallel propagating WHFI. In Section 3 we obtain the marginal stability thresholds of the WHFI for different values of $T_{\parallel s}/T_{\parallel c}$ and $n_s/n_e$ and present the best fit parameters for easier comparison with observational data. Finally, in Section 4 we present the summary and conclusions of this work. 

\section{WHFI dispersion relation in the context of the core-strahlo Model}

To study the excitation of the parallel propagating whistler mode in a solar-wind-like plasma, we describe the electron population using the core-strahlo model. As already mentioned, this model was first proposed as a eVDF for the solar wind in paper A, where the authors showed that it is able to reproduce the quasi-thermal core, high energy tails, and field-aligned skewness observed in the eVDF. In this description, the electron distribution $f_e$ is given by Eq. (\ref{eq_total_dist}) and consist on a superposition of a quasithermal core $f_c$, described by a drifting biMaxwellian; and a suprathermal strahlo $f_s$, described by a Skew-Kappa function. 
\begin{equation}
\label{eq_total_dist}
f_e(v_{\perp}, v_{\parallel}) = f_c(v_{\perp}, v_{\parallel})+ f_{s}(v_{\perp}, v_{\parallel}),    
\end{equation}
where
\begin{equation}
\label{eq_core_dist}
    f_c(v_{\perp}, v_{\parallel}) = \frac{n_c}{\pi^{3/2}\alpha_{\perp}^2\alpha_{\parallel}} \ \exp\left(-\frac{v_{\perp}^2}{\alpha_{\perp}^2} - \frac{(v_{\parallel}-U_{c})^2}{\alpha_{\parallel}^2}\right),
\end{equation}
and
\begin{equation}
    \label{eq_skd}
f_{s}(v_{\perp}, v_{\parallel}) = n_{s}A_{s}\left[1+ \frac{1}{\kappa_{s} - \frac{3}{2}}\left(\frac{v_{\bot}^2}{\theta_{\bot}^2} + \frac{v_{\parallel}^2}{\theta_{\parallel}^2} + \delta_{s}\left( \frac{v_{\parallel}}{\theta_{\parallel}} - \frac{v_{\parallel}^3}{3\theta_{\parallel}^3}\right)\right)\right]^{-(\kappa_{s}+1)}\,.
\end{equation}
In the above expressions the sub-indexes $\parallel$ and $\bot$ are with respect to the background magnetic field, $n_c$ and $n_s$ represent the core and strahlo number density, $\alpha_\parallel$ and $\alpha_\bot$ correspond to the thermal velocities of the core subpopulation, $U_c$ is the core drift velocity, and $\theta_\parallel$ and $\theta_\bot$ are related to the thermal velocities of the strahlo. Additionally, the skewness parameter $\delta_s$ modifies the field-aligned skewness such that higher $\delta_s$ values indicate more skewed distributions. Furthermore, the kappa parameter $\kappa_s$ controls the slope of the high energy tails such that as $\kappa_s$ increases, the enhanced tails of distribution Eq. (\ref{eq_total_dist}) diminish (see figure 3 in paper A). In paper A, the authors examined the behavior of the Skew-Kappa distribution Eq. (\ref{eq_skd}) in velocity space, which allowed them to establish a validity range for the core-strahlo model. Accordingly, we must impose small skweness i.e. $\delta_s^3 \ll 1$ for this description to be applicable as a distribution function for the solar wind's electrons. Moreover, the core-strahlo model must fulfill the quasi-neutrality condition:
\begin{equation}
    \label{eq_qn}
    n_c + n_s = n_e = n_p,
\end{equation}
and also be current-free (see paper A for details). 
\begin{equation}
    U_{c}\ =\  \frac{n_{s}}{n_c}\ \frac{\delta_s}{4}\theta_{\parallel}.
    \label{eq_Uc}
\end{equation}

\begin{figure}[ht]
\begin{center}
\includegraphics[width=14cm]{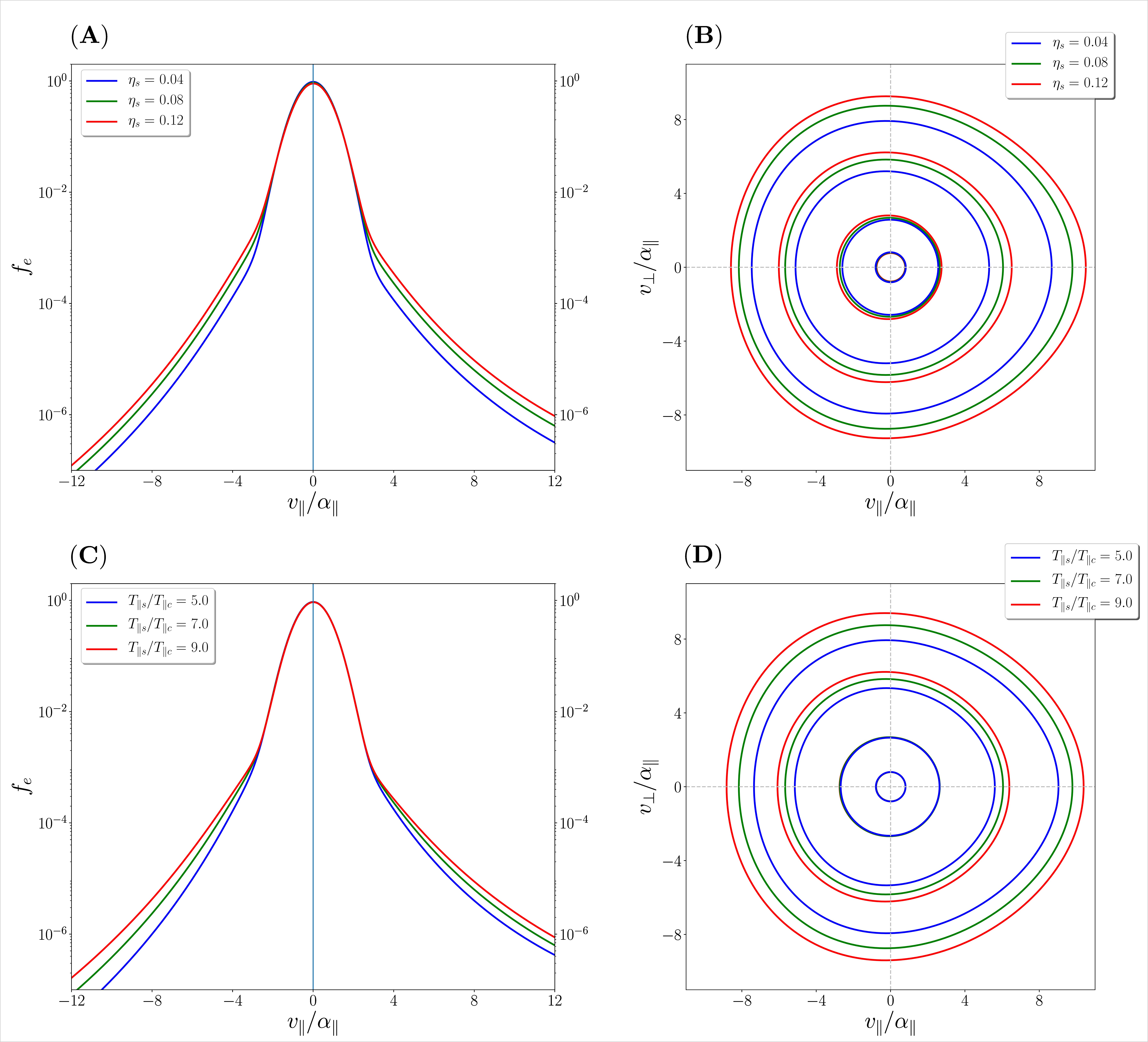}
\end{center}
			\caption{Parallel cuts (left) and contour plots (right) of the core-strahlo distribution from Eq.~\ref{eq_total_dist}. Top panels consider fixed $T_{\parallel s}/T_{\parallel c} =7.0$, and different densities $\eta_s=0.04$ (blue), $\eta_s=0.08$ (green), and $\eta_s=0.12$ (red); Bottom panels consider fixed density ($\eta_s=0.08$), and different temperature ratios $T_{\parallel s}/T_{\parallel c} =5.0$ (blue), $T_{\parallel s}/T_{\parallel c} =7.0$ (green), and $T_{\parallel s}/T_{\parallel c} =7.0$ (red). In all panels, we considered a skewness parameter $\delta_s = 0.2$, a kappa parameter $\kappa_s=3.0$ and set the anisotropy for the electron core and strahlo equal to one.}
\label{fig_dist}
\end{figure}

As we previously pointed out, it has been reported in several works that the values of the relative density of the non-thermal electron population (the strahlo in this representation) and the temperature ratio between different subpopulations vary throughout the Heliosphere. Thus, it becomes relevant to understand how the total eVDF modifies with these parameters and the impact these changes have on the WHFI. In addition, $T_{\parallel s}/T_{\parallel c}$ and $\eta_s = n_s/n_e$ are the last two parameters that determine the shape of distribution Eq. (\ref{eq_total_dist}) that remains to be analyzed in the isotropic case ($\alpha_\bot=\alpha_\parallel$, and $\theta_\bot = \theta_\parallel)$. Accordingly, Figure \ref{fig_dist} shows parallel cuts at $v_\perp =0$ (left panels) and contour plots (right panels) of the core-strahlo distribution for: fixed $T_{\parallel s}/T_{\parallel c} = 7.0$ and different values the relative density of the strahlo subpopulation $\eta_s =\ 0.04,\ 0.08,\ 0.12 $ (top panels); and fixed $\eta_s = 0.08$ with different values of the strahlo-to-core parallel temperature ratio $T_{\parallel s}/T_{\parallel c} =  5.0,\ 7.0,\ 9.0$ (bottom panels). To obtain all these plots, we consider isotropic core and strahlo distributions, with a skewness parameter $\delta_s = 0.2$, and a kappa parameter $\kappa_s = 3.0$. We can notice in this figure that the core-strahlo distribution display field-aligned skewness, enhanced tails and a narrower Maxwellian core, as was already established. From Panels \ref{fig_dist}A and \ref{fig_dist}B, we can see that the relative density of the strahlo subpopulation modifies the high energy tail of the distribution so that the tails of the distribution are enhanced, as the Skew-Kappa function describing the strahlo goes up, with increasing $\eta_s$. We can also notice that changes in $\eta_s$ have a minor effect on the quasithermal core of the eVDF. Namely, as $\eta_s$ increases the core has a slight decrease in amplitude.

Moreover, from panel \ref{fig_dist}A it seems that the slope of these energetic tails is not altered with $\eta_s$. The field-aligned skewness of the core-strahlo distribution appears to remain unchanged as well, which is more evident in the contour plot shown in panel \ref{fig_dist}B (compared with figure 3b on paper A). We encounter a similar behavior when we modify the strahlo-to-core temperature ratio $T_{\parallel s}/T_{\parallel c}$, as we can see in panels \ref{fig_dist}C and \ref{fig_dist}D. It is clear that this parameter also modifies the high energy tails of the core-strahlo distribution, as the Skew-Kappa function describing the strahlo subpopulation widens with increasing $T_{\parallel s}/T_{\parallel c}$. We can also see that for higher values of $T_{\parallel s}/T_{\parallel c}$, the energetic tails are enhanced but, unlike the previous parameter, the Maxwellian core appears to remain the same. Moreover, it seems that the skewness of the core-strahlo distribution does not change when $T_{\parallel s}/T_{\parallel c}$ is modified, which is noticeable in the contour plot shown in panel \ref{fig_dist}D. Accordingly, both parameters, the strahlo-to-core temperature ratio and the density of the strahlo subpopulation, can alter the tails of distribution \ref{eq_total_dist}. The general behavior is that as $T_{\parallel s}/T_{\parallel c}$ and $\eta_s$ decrease, the high energy tails diminish, while maintaining the skewness of the distribution unaltered. For both parameters we have use representative values that have been measured in the solar wind at different solar distances \citep{pierrard2016electron,lazar2020characteristics}. This dependence of the core-strahlo distribution on $T_{\parallel s}/T_{\parallel c}$ and $\eta_s$ may influence the excitation of the WHFI, which we will study next. It is worth mentioning that it is the field-aligned skweness the non-thermal feature that provides the free energy for the excitation of the WHFI and, in this representation, it can be modified mostly through the skewness parameter $\delta_s$. Nevertheless, as $T_{\parallel s}/T_{\parallel c}$ and $\eta_s$ can also regulate the shape of the eVDF, and there is wide evidence that these parameters have several values throughout the Heliosphere, here we focus our analysis on the effect of them on the WHFI.

\subsection{WHFI dispersion relation}
\label{WHFI_DR}

Now we focus on the stability of the WHFI and study the effect $T_{\parallel s}/T_{\parallel c}$ and $\eta_s$ have on the dispersion relation of the parallel propagating whistler mode. We model the solar wind populations using the core-strahlo distribution (\ref{eq_total_dist}) for the electrons and a Maxwellian function for the protons. The procedure to obtain the dispersion relation for wave modes that propagate in this system parallel to the background magnetic field $\vec{B}_0=B_0\hat{z}$ such that $\vec{k}=k\hat{z}$ was already discussed in detail in paper A, where an analytical expression for the dispersion tensor in the validity range of the model (i.e $\delta_s^3 \ll 1$) can be found (see Appendix B in paper A). The dispersion relation $\omega = \omega(k)$ between the wavenumber $k$ and the complex wave frequency $\omega = \omega_r + i \gamma$ for the parallel propagating WHFI is obtained numerically in this analysis. We consider a proton population such that $\beta_{\parallel p} =0.1$ where $\beta_{\parallel j}$ is the plasma beta of population $j$. For the eVDF we again set the kappa parameter to $\kappa_s = 3.0$, the skewness parameter to $\delta_s = 0.2$ and work with isotropic subpopulations such that $T_{\bot c}/T_{\parallel c} = 1.0$ and $T_{\bot s}/T_{\parallel s} = 1.0$, We also fix the strength of the background magnetic field so that $\beta_{\parallel s} = 1.0$ and set the ratio between the electron plasma frequency ($\omega_{pe})$ and electron gyrofrequency $(\Omega_e)$ to $\omega_{pe}/|\Omega_e|= 200$.
Hence, with this selection of parameters, the only relevant non-thermal features in the study are the high energy tails and field-aligned skewness.
Lastly, to analyze how the excitation of the whistler mode depends on $\eta_s$ and $T_{\parallel s}/T_{\parallel c}$, we use values between $T_{\parallel s}/T_{\parallel c} = 3.0$ and $T_{\parallel s}/T_{\parallel c}=11.0$ for the strahlo-to-core parallel temperature ratio and relative density for the strahlo up to 12\% (i.e $\eta_s=0.12$), all of which have been measured in the solar wind as reported by \citet{lazar2020characteristics}.

\begin{figure}[ht]
\begin{center}
\includegraphics[width=17cm]{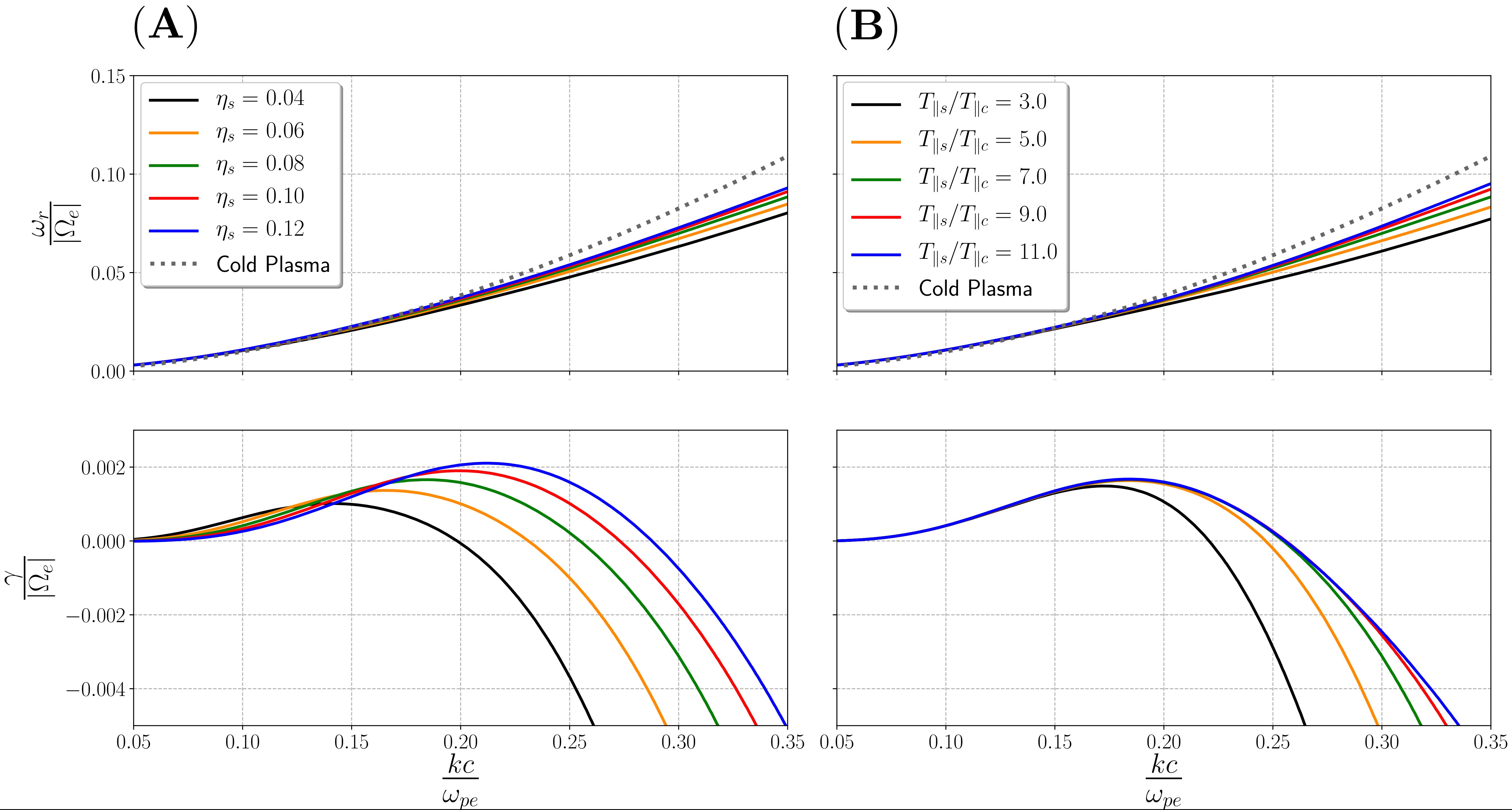}
\end{center}
			\caption{Real (top) and and imaginary (bottom) parts of the dispersion relation $\omega=\omega(k)$ for the whistler mode. Panel (A) consider $T_{\parallel s}/T_{\parallel c}=7.0$ and different $\eta_s$. Panel (B) consider $\eta_s = 0.08$ and different $T_{\parallel s}/T_{\parallel c}$ values. For these plots, we set $\beta_{\parallel s} = 1.0$ and all other parameters are the same as in Figure \ref{fig_dist}.}
\label{fig_disp_rel}
\end{figure}

Figure \ref{fig_disp_rel} shows the dispersion relation of the parallel-propagating whistler mode for: fixed $T_{\parallel s}/T_{\parallel c} = 7.0$ and different values of $\eta_s$ (left panel) and fixed $\eta_s = 0.08$ and different $T_{\parallel s}/T_{\parallel c}$ values (right panel). Top and bottom panels show, respectively, the real frequency $\omega_r$ and imaginary frequency $\gamma$, both expressed in units of $|\Omega_e|$ and as a function of the normalized wavenumber $kc/\omega_{pe}$, where $c/\omega_{pe}$ is the electron inertial length. From Panel \ref{fig_disp_rel}A we can see that, in the wavenumber range shown, the real part of the frequency has weak dependence on $\eta_s$ such that $\omega_r$ slightly increases with this parameter. For comparison purposes we also include the cold plasma dispersion relation. We can see that for the case of the cold dispersion the frequency is larger, which is expected and consistent with previous studies~\citep[see e.g.][]{kuzichev2019nonlinear}. From the imaginary part of the frequency, $\gamma$, we can see that the waves become more unstable as the strahlo relative density increases: the wavenumber range in which $\gamma >0$ widens and the maximum growth rate value $\gamma_{\rm{max}}$ for this mode increases with $\eta_s$. Considering that the strahlo is the subpopulation that provides the free energy to radiate, it is expected the plasma to become more unstable with increasing $\eta_s$, as a higher value of this parameter represents a more important non-thermal subpopulation relative to the core, as shown in Figure \ref{fig_dist}A. This is consistent with similar already reported results but based on a model composed by two drifting Maxwellian VDFs (see for example figure 3 in \citet{gary1985electromagnetic}). 

On the other hand, from panel \ref{fig_disp_rel}B we can see that the real part of the frequency decreases when the strahlo-to-core temperature ratio decreases. The imaginary part $\gamma$, however, does not have such a straightforward behavior. 
For lower values of $T_{\parallel s}/T_{\parallel c}$, the wave mode becomes more unstable as this parameter increases, which is noticeable for the solutions with $T_{\parallel s}/T_{\parallel c} = 3.0$ and $T_{\parallel s}/T_{\parallel c} = 5.0$ (black and yellow curves, respectively). The wavenumber range in which the growth rates are positive widens and $\gamma_{\rm{max}}$ slightly increases with increasing $T_{\parallel s}/T_{\parallel c}$. From $T_{\parallel s}/T_{\parallel c} = 5.0$ onward, however, the changes in $\gamma$ with temperature ratio are barely noticeable and the curves remain almost the same. This behavior is maintained for even higher $T_{\parallel s}/T_{\parallel c}$ values than those shown in this plot. Therefore, a higher temperature (with respect to the core) of the subpopulation that provides the free energy (i.e the strahlo), does not further destabilize the plasma above the saturation point $T_{\parallel s}/T_{\parallel c} \approx 5.0$. A similar result can be seen in Figure 5b of paper A, where the growth rates also saturate at $T_{\parallel s}/T_{\parallel c} \approx 5.0$ for other plasma parameters. The complete characterization of the saturation point seems interesting, especially when considering that other models do not present this feature (see figure 4 in \citet{gary1985electromagnetic}, for example) but it requires a more in-depth analysis, beyond the scope of this study. 

In summary, as we are using values relevant for the solar wind plasma, it is important to emphasize that the changes introduced by $T_{\parallel s}/T_{\parallel c}$ and $\eta_s$ on the stability of the parallel propagating whistler mode, regardless how weak they seem), may have an impact on the thresholds we use to compare with observational data. This may be relevant to assess the importance of the WHFI, and the relative importance of its marginal stability thresholds, in the non-collisional regulation of the electron heat-flux in the solar wind.

\section{WHFI $\beta$ and $\delta$ thresholds}

In this section, we systematize the linear analysis of the parallel propagating WHFI and present the marginal stability thresholds for this mode as function of $\delta_s$ and $\beta_{\parallel c}$. To obtain these thresholds, we describe the plasma populations as in the previous section. We use a Maxwellian distribution to model the protons and set $\beta_{\parallel p} = 0.1$. For the electrons we use the core-strahlo distribution with isotropic subpopulations and set $\kappa_s = 3.0$. Thus, with all these fixed parameters, we calculate the normalized maximum growth rate $\gamma_{\rm{max}}/|\Omega_e|$ of the parallel propagating whistler mode in the $\delta_s -\beta_{\parallel c}$ space. We repeat this procedure for different values of $\eta_s$ and $T_{\parallel s}/T_{\parallel c}$ (as previously reported using solar wind electron measurements) to understand under which plasma conditions the whistler mode destabilize in the context of the core-strahlo model. Following the suggestion proposed in paper A, here we are presenting the stability thresholds as a function of the skewness parameter $\delta_s$ instead of the electron heat-flux macroscopic parameter $q_{\parallel e}$, which has been customarily used for this purpose in the past. As $q_{\parallel e}$ is a moment of the distribution function, its expression depends on all the parameters that determine its shape in velocity space. In our case, the analytical expression for the electron heat-flux in the validity range of the core-strahlo model, when considering isotropic subpopulations, is given by
\begin{equation}
\label{eq_norm_hf}
		\dfrac{q_{\parallel e}}{q_0} = \frac{\delta_s}{3}\frac{n_s}{n_e}\left(\frac{T_{\parallel s}}{T_{\parallel c}}\right)^{\frac{3}{2}}\left[\frac{7}{4\kappa_s -10}+\frac{5}{4}\frac{T_{\parallel c}}{T_{\parallel s}} - \frac{3}{4}\right],
\end{equation}
where $q_0$ is the free-streaming heat-flux (see paper A for details). As shown in Eq. (\ref{eq_norm_hf}), for fixed $\kappa_s$ the heat-flux moment depends on all relevant parameters, and can have the same value for different combinations between them. Thus, to avoid this issue we solve the dispersion relation in the $\delta_s -\beta_{\parallel c}$ space. Nevertheless, using Eq. (\ref{eq_norm_hf}) it is not difficult to find the same thresholds in terms of $q_{\parallel}e/q_0$ and $\beta_{\parallel c}$. 

\begin{figure}[ht]
\begin{center}
\includegraphics[width=17cm]{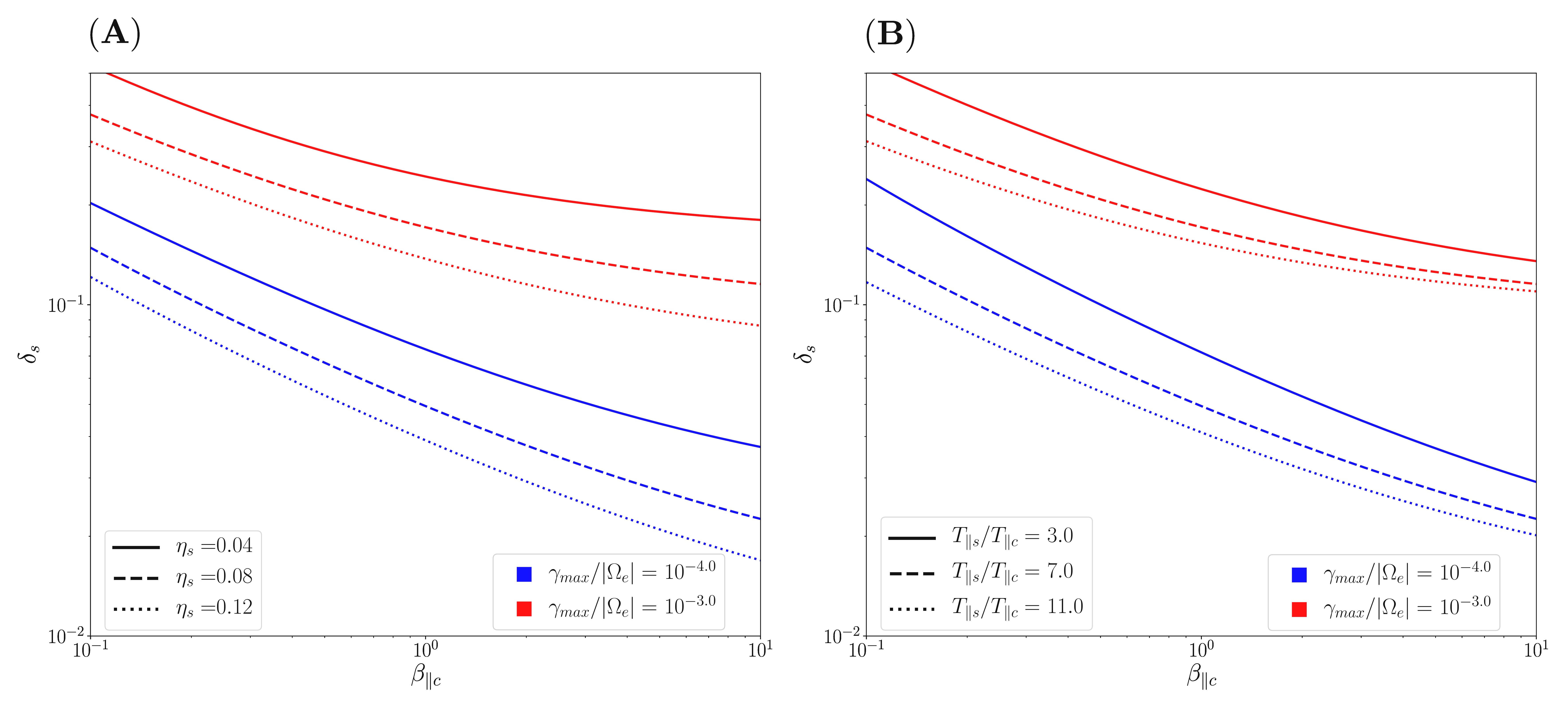}
\end{center}
\caption{Instability thresholds  $\gamma_{\rm{max}}/|\Omega_e| = 10^{-3}$ (red lines) and $\gamma_{\rm{max}}/|\Omega_e| = 10^{-4}$ (blue lines) of the whistler heat-flux instability for: (A) $T_{\parallel s}/T_{\parallel c}=7.0$ and different $\eta_s$ values; (B) $\eta_s = 0.08$ and different $T_{\parallel s}/T_{\parallel c}$ values. All calculations were performed using isotropic subpopulations and $\kappa_s = 3.0$.}
\label{fig_thresholds}
\end{figure}

Figure \ref{fig_thresholds} shows the contour levels $\gamma_{\rm{max}}/|\Omega_e| = 10^{-3}$ (red lines) and $\gamma_{\rm{max}}/|\Omega_e| = 10^{-4}$ (blue lines) of the normalized maximum growth rate for different values of $\eta_s$ (left panel) and $T_{\parallel s}/T_{\parallel c}$ (right panel).  
Panel \ref{fig_thresholds}A shows these thresholds for a fixed value of $T_{\parallel s}/T_{\parallel c} = 7.0$, and $\eta_s=0.04, 0.08$ and $0.12$ (solid, dashed and pointed lines, respectively). We can see the thresholds move down and to the left as we increase the strahlo relative density. As expected, the plasma is more easily destabilize for higher values of $\eta_s$. In other words, as we increase $\eta_s$, lower values of of $\delta_s$ or $\beta_{\parallel c}$ are needed to produce the same growth rate of the WHFI. On the other hand, Panel \ref{fig_thresholds}B shows the contours for $T_{\parallel s}/T_{\parallel c} = 3.0, 7.0$ and $11.0$ (solid, dashed and pointed lines, respectively) and fixed $\eta_s = 0.08$. We can see the same trend as in the previous plot. The plasma becomes more unstable to the parallel propagating WHFI as  $T_{\parallel s}/T_{\parallel c}$ increases, so that the thresholds move to the left and downward. 

Finally, to facilitate the comparison between observational data and the linear prediction for the stability of the WHFI, we fit the contour lines $\gamma_{\rm{max}}/|\Omega_e| = 10^{-3}$ and $10^{-4}$ using a generalized Lorentzian function; namely
\begin{equation}
	\label{eq_fit}
		\delta_s = A + \frac{B}{\left(\beta_{\parallel c} - \epsilon_0^2\right)^\alpha}.
\end{equation}
The best-fit value for parameters $A$, $B$, $\epsilon_0$ and $\alpha$ of every threshold shown in Figure \ref{fig_thresholds} can be seen in Table \ref{table_fits} for direct comparison with data in the beta range shown. With the results shown for $\eta_s$ and $T_{\parallel s}/T_{\parallel c}$, we have established that each of them modify the stability of the whistler mode in a distinct way and with different strength. This reinforce the conclusion that it is not possible to assess if a plasma state is stable to the WHFI through $q_{\parallel e}$ without having additional information about the shape of the distribution and its dependence on all plasma parameters. Therefore, we believe that the role of WHFI in the relaxation process of plasma states should be studied in terms of microscopic parameters that determine the eVDF and not only macroscopic moments.

\begin{table}[h!]
	\caption{Best fit parameters for the $\gamma_{\rm{max}}/|\Omega_e| =  10^{-3}$ and $10^{-4}$ thresholds of the whistler heat-flux instability. The curve fitting for these thresholds was performed using the function shown in eq. (\ref{eq_fit}}
	\begin{tabular}{c|c|c||c|c|c|c|}
				\multicolumn{3}{c||}{} & $A$ & $B$ & $\epsilon_0$ & $\alpha$ \\				
				\hline \hline
			    & $\eta_s = 0.04$ & $T_{\parallel s}/T_{\parallel c}=7.0$ & 0.162 & 0.082 & $2.4 \times 10^{-4}$ & 0.648 \\ 
				\cline{2-7}
				& & $T_{\parallel s}/T_{\parallel c} = 3.0$ & 0.102 & 0.122 & $4.9\times 10^{-6}$ & 0.559 \\ 
				\cline{3-7}

				$\gamma_{\rm{max}} = 10^{-3}$& $\eta_s = 0.08$ & $T_{\parallel s}/T_{\parallel c} =7.0$ & 0.094 & 0.077& $4.8 \times 10^{-6}$ & 0.562 \\ 
				\cline{3-7}

				& & $T_{\parallel s}/T_{\parallel c} = 11.0$ & 0.093 & 0.061 & $7.1 \times 10^{-5}$ & 0.557 \\ 
				\cline{2-7}
				
				& $\eta_s = 0.12$ & $T_{\parallel s}/T_{\parallel c}=7.0$ & 0.065 & 0.073 & $3.2 \times 10^{-6}$ & $0.530$ \\ 
				\hline \hline

			    & $\eta_s = 0.04$ & $T_{\parallel s}/T_{\parallel c}=7.0$ & 0.023 & 0.050 & $1.5 \times 10^{-5}$ & 0.554 \\ 
				\cline{2-7}
				& & $T_{\parallel s}/T_{\parallel c}=3.0$ & 0.013 & 0.059 & 0.117 & 0.553 \\ 
				\cline{3-7}

				$\gamma_{\rm{max}} = 10^{-4}$& $\eta_s = 0.08$ & $T_{\parallel s}/T_{\parallel c} = 7.0$ & 0.012 & 0.037 & 0.094 & 0.541\\ 
				\cline{3-7}

				& & $T_{\parallel s}/T_{\parallel c} = 11.0$ & 0.012 & 0.029 & 0.076 & 0.540 \\ 
				\cline{2-7}
				
				& $\eta_s = 0.12$ & $T_{\parallel s}/T_{\parallel c}=7.0$ & 0.008 & 0.031 & 0.107 & 0.538 \\ 
				\hline \hline				
			\end{tabular}
			\label{table_fits}
	\end{table}

\section{Summary and Conclusions}

In this work we have used the core-strahlo model to describe the eVDF in the solar wind, and analyzed the stability of the parallel propagating whistler mode in a magnetized non collisional plasma. We have shown how the electron distribution modifies with the strahlo relative density and the strahlo-to-core temperature ratio density, and the impact these changes have on the excitation of the WHFI, as well as in the stability thresholds in delta-beta space. The general behavior is that as $\eta_s$ and $T_{\parallel s}/T_{\parallel c}$ increase, the plasma becomes more unstable to the WHFI. However, the dependence on $T_{\parallel s}/T_{\parallel c}$ is much weaker, and above certain level ($T_{\parallel s}/T_{\parallel c}\sim 5$) the changes in growth rates are no longer noticeable. We have also shown the enhancing effect of $\eta_s$ and $T_{\parallel s}/T_{\parallel c}$ on the stability thresholds in delta-beta space and provided the best-fit parameters for comparison with observations. With these results we have studied the dependence of the stability of the whistler mode on all the parameters that determine the shape of the eVDF in the isotropic case. Therefore, the usage of the core-strahlo model allowed us to study the WHFI in all the relevant parameter space in a manageable way, but considering a realistic representation of the solar wind electron population, including quasi-thermal core, high energy tails, and field-aligned skewness in the analysis all at once. It is important to mention that, besides the asymmetry represented by the heat-flux, temperature anisotropy should also play a role. However, as shown by several studies, among the anisotropic states, the isotropic state is also ubiquitous to the solar wind at different solar distances and solar wind speeds \citep[see e.g.][]{Adrian2016,lazar2020characteristics}, and here we have focused on the effect of asymmetry by itself. A systematic study on the combined effect of both free energy sources (asymmetry and anisotropy), and the subsequent interplay between the WHFI and electron-cyclotron or firehose instabilities should be also relevant but is beyond the scope of this study. 

As mentioned, it has been reported in several works that the parameters here studied change with radial distance from the Sun \citep{pierrard2016electron,lazar2020characteristics,cattell2022parker}. For example, in \citet{lazar2020characteristics}, the authors showed that the average temperature ratio between the halo and core subpopulations varies from $T_h/T_c \sim 8$ at 1 AU, to $T_h/T_c \sim 3$ at 0.3 AU. A variation with solar wind conditions (slow and fast wind) also exists, such that even at a given radial distance the measurements vary considerably, ranging between $T_h/T_c \sim 2$ and $T_h/T_c \sim 15$ at 1 AU for the temperature ratio, and between less that 1\% up to 15\% for the relative density of the halo (see Figure 2 and 3 in \citet{lazar2020characteristics}). Accordingly, we believe that efforts should be made, to take into consideration the real impact that these parameters have on the stability of the WHFI. This should be particularly relevant when assessing the WHFI relevance on the non-collisional regulation of the heat-flux through comparison between theoretical prediction and data. We expect these predictions to be assessed and validated with electron measurements obtained with current and new solar wind missions. Systematic theoretical studies considering realistic solar wind conditions, and also comparisons between the results obtained with different kinetic model of the solar wind plasma, may be relevant in order to adequately understand the heat-flux transport through the Heliosphere.

\section*{Conflict of Interest Statement}
The authors declare that the research was conducted in the absence of any commercial or financial relationships that could be construed as a potential conflict of interest.

\section*{Author Contributions}
All authors listed have made a substantial, direct and intellectual contribution to the work, and approved it for publication. All authors contributed equally to this work.

\section*{Funding}
This research was funded by ANID, Chile through the Doctoral National Scholarship N$^{\circ}$21181965 (B.Z.Q.) and FONDECyT grant No. 1191351 (P.S.M.).

\section*{Acknowledgments}
We would like to thank Dr. Adolfo F. Viñas for useful discussion. We also gratefully acknowledge the support by ANID, Chile through a Doctoral Scholarship and a Fondecyt grant.

\section*{Data Availability Statement}
The original contributions presented in the study are included
in the article/supplementary material, further inquiries can be
directed to the corresponding authors.

\bibliographystyle{frontiersinSCNS_ENG_HUMS} 
\bibliography{manuscript.bib}





\end{document}